\documentclass[aps,twocolumn,a4paper,pra,superscriptaddress,floatfix]{revtex4}%
\usepackage{amssymb}
\usepackage{amsmath}
\usepackage{amsfonts}
\usepackage{graphicx}
\usepackage{graphics}
\usepackage{color}

\def\lsim{\mathrel{\rlap{\lower4pt\hbox{\hskip1pt$\sim$}}
    \raise1pt\hbox{$<$}}}
\def\gsim{\mathrel{\rlap{\lower4pt\hbox{\hskip1pt$\sim$}}
    \raise1pt\hbox{$>$}}}

\def\bea{\begin{eqnarray}}
\def\bee{\end{eqnarray}}

\begin{document}

\title{Pion form factors} 

\author{Mikhail Gorchtein} 
\affiliation{Center of Exploration of Energy and Matter, Indiana University, Bloomington, IN 47408} 
\author{Peng Guo}
\affiliation{Physics Department, Indiana University, Bloomington, IN 47405} 
\author{Adam P. Szczepaniak}
\affiliation{Physics Department, Indiana University, Bloomington, IN 47405} 
\affiliation{Center of Exploration of Energy and Matter, Indiana University, Bloomington, IN 47408} 

\date{\today}

\begin{abstract}
We consider the electromagnetic and transition pion form factors. 
Using dispersion relations we simultaneously describe both the hadronic, time-like region and the asymptotic region of large energy-momentum transfer. 
For the latter we propose a novel mechanism  of Regge fermion exchange. We find that hadronic contributions dominate form factors at all currently available energies. 
\end{abstract}


\maketitle

Photons interact with quarks,  the charged constituents of hadrons and the resulting electromagnetic form factors probe the quark energy-momentum distribution in hadrons. 
 In this letter we  examine  the charged pion electromagnetic form
 factor $F_{2\pi}(s)$, which is   defined  by the matrix element  
$\langle \pi^+(p') \pi^-(p) |J_\mu |0 \rangle = e(p' -p)_\mu F_{2\pi}(s)$,  
 and the transition from factor 
between the neutral pion and a real photon, $F_{\pi\gamma}(s)$ determined  by $\langle \pi^0(p')\gamma(\lambda, p) |J_\mu |0 \rangle = 
 ie^2/4\pi^2 f_\pi  \epsilon_{\mu \nu \alpha\beta} \epsilon^{*\nu}(\lambda) p'^\alpha p^\beta  F_{\pi\gamma}(s) $. 
 Above, $J_\mu$ is the electromagnetic current,  $s = (p' + p)^2$ is
 the four-momentum transfer squared and $f_\pi = 92.4\mbox{ MeV}$ 
is the pion decay constant. Current conservation implies 
 $F_{2\pi}(0) = 1$ and, in the chiral limit, axial
 anomaly determination of the  $\pi^0 \to 2\gamma$ decay leads to the expectation $F_{\pi\gamma}(0) \approx 1$.  Because  at short distances quark/gluon interactions are asymptotically free, it has been postulated  that at high energy or momentum transfer,  $|s| \gg \mu^2$, both form factors measure 
   hard scattering  of the photon with a small number of the QCD constituents~\cite{Efremov:1978rn,Duncan:1979hi,brodsky-lepage}.
One would then expect $\mu^2\sim O(1\mbox{ GeV}^2)$,  which is
 the typical hadronic scale,  however,  given the  current  status of the  data  it  seems  that
$\mu^2$ could be as large as 
$ ~O(10 - 100\mbox{GeV}^2)$~\cite{Szczepaniak:1997sa,Efremov:2009dx}. 
This implies that an alternative description of the underlying dynamics might  be in order and 
   the subject of   applicability of pQCD to exclusive reactions has in fact a  long history~\cite{Isgur:1988iw}. 
 Perturbative QCD (pQCD) analysis of the form factor
asymptotics  assumes specific properties of  certain
non-perturbative quantities, {\it i.e.}  the parton momentum 
distribution amplitudes in the low-momentum,''wee" region. If these had different behavior 
 from what is assumed in the pQCD analysis  
 the arguments leading to dominance of leading twist perturbative
scattering would  break down~\cite{Sucher:1971kx}. 
Such pion distribution amplitudes were considered recently in ~\cite{RuizArriola:2002bp,Dorokhov:2006qm}, however, the authors used perturbative evolution to soften the wee region and use pQCD formulae.  The available data
on the pion electromagnetic form factor ranges up to 
$|s|  \lsim 10 \mbox{ GeV}$ \cite{fpi_jlab}
and is approximately a factor of three above 
the asymptotic prediction \cite{brodsky-farrar}.  
  Even more spectacular discrepancy is observed  in  the  transition form factor
  recently measured by  BaBar~\cite{babar}.  
 For momentum transfers 
as  large as  $-s \approx 40 \mbox{GeV}^2$ the measurement disagrees with the asymptotic prediction not only in 
  normalization but also in the overall $s$-dependence. While pQCD
  predicts $ s F_{\pi\gamma}(s) \to 2 f_\pi$  as $|s| \to
  \infty$ \cite{brodsky-lepage},  the data suggest that the magnitude
  of  $-s F_{\pi\gamma}(s) $ grows with $|s|$.
 
Crossing symmetry implies that form factors in the  space-like ($s < 0$) and time-like ($s > 0$)  
 region are boundary values of  an analytical function defined in the
 complex-$s$ plane  with a unitarity cut 
  running over the positive $s$-axis and starting at the  two pion production threshold branch point  $s_{th} = 4m_\pi^2$.  
  In the time-like region the electromagnetic (transition) form factor describes the amplitude for 
   production  of a  spin-1, $\pi^+\pi^-$ pair   ($\pi^0\gamma$)  
   in the external  electromagnetic field of the virtual photon. 
 In the space-like region the form factors are usually interpreted in terms of 
  parton  three-momentum distribution in a hadron (and/or photon). 
Analyticity demands these  apparently distinct  physical pictures to be smoothly connected.  
The dominant feature of the 
   spin-$1$,  $\pi^+\pi^-$  state is the isovector $\rho(770)$ resonance, which 
    also dominates the electromagnetic form factor. There is no
    time-like data 
 available  for the transition form factor, however, also  in this case one
 expects to see the $\rho$, and  the isoscalar, $\omega(782)$ 
 resonance. The analytical continuation to the space-region implies 
 that for $-s \lsim 1\mbox{GeV}^2$, {\it i.e.} in the hadronic range, 
   the quark wave function is dual to the  vector-meson exchange in the crossed channel.  

In the following, we relate the
space-like and time-like regions through  dispersion relations, 
 and focus  on the  dynamics in the asymptotic region, $s
\to + \infty$. 
In view of the BaBar "anomaly"  and the apparent failure of the pQCD 
description, we  propose a novel description 
 for the  dominant mechanism that drives the asymptotic behavior of the form factors. 

The discontinuity   of
  $F_{\pi\gamma}(s)$ across the unitary  cut  is given by 
\begin{equation} 
Im F_{\pi\gamma} = t^*_{2\pi,\pi\gamma} \rho_{2\pi} F_{2\pi} +  t^*_{3\pi,\pi\gamma} \rho_{3\pi}  F_{3\pi} + 
\sum_{X} t^*_{X,\pi\gamma} \rho_X F_X  \label{imt}
\end{equation} 
 and the sum runs over all possible intermediate states $X\neq2\pi,3\pi$.
 Here,  $t_{X,\pi\gamma}$ ($F_{X}$) represent the amplitudes for 
$X\to \pi^0 \gamma$ ($\gamma^* \to X$), respectively and 
$\rho_{X}$ is a product of the phase space and kinematical factors 
  ({\it i.e.}  for the $2\pi$ intermediate state  $\rho_{2\pi}(s) = s (1- s_{th}/s)^{3/2}/96\pi$).
 Provided Im$\,F_{\pi\gamma}$ vanishes at $s\to\infty$, its real
 part can be reconstructed for any $s$ from the unsubtracted dispersion relation 
  \begin{equation} 
 F_{\pi\gamma}(s) = \frac{1}{\pi} \int_{s_{th}}  ds' \frac{Im
   F_{\pi\gamma}(s')}{s' - s}. 
\label{disp}
 \end{equation} 
The two lowest mass intermediate states, $X=2\pi,\,3\pi$ that are dominated 
  by the $\rho(770)$ and $\omega(782)$ resonances, respectively, are expected 
to saturate the cut in the hadronic range $s_{th} < s  \lsim  1 \mbox{GeV}^2$.  
The $\omega(782)$ in the isoscalar $3\pi$ channel 
is a narrow  resonance with width to mass ratio,   $\Gamma_\omega/m_\omega \sim10^{-2}$ 
and its contribution to $F_{\pi\gamma}$ can be well approximated by a 
Breit-Wigner distribution,
 \begin{equation} 
 F_{\pi\gamma}^{(3\pi)}(s)  =  
\frac{ c^{(3\pi)}_{\pi\gamma} m_\omega^2}
{m_\omega^2 - s - i m_\omega \Gamma_\omega(s)}  
\label{omega} 
\end{equation}  
with $c^{(3\pi)}_{\pi\gamma}   = 4\pi^2 f_\pi g_{\omega\pi\gamma}/ m_\omega g_\omega = 
      0.493 $ obtained from $\omega \to \pi\gamma$ 
and $\omega \to e^+e^-$ decay widths yielding 
       $g_{\omega\pi\gamma} = 1.81$ and $g_\omega = 17.1$,
       respectively. 
The contribution from the $2\pi$ intermediate state is 
 dominated  by the $\rho(770)$ resonance, which determines both the
$t_{2\pi,\pi\gamma}$ 
 scattering amplitude and the pion electromagnetic form factor, 
$F_{2\pi}$ for  $s  \lsim 1 \mbox{GeV}^2$, 
and vector meson dominance (VMD), yields  
$c^{(2\pi)}_{\pi\gamma}   = 4\pi^2 f_\pi g_{\rho\pi\gamma}/ m_\rho g_\rho = 
      0.613 $ 
(with $\rho \to \pi\gamma$ and $\rho \to e^+e^-$ decay widths leading to  
       $g_{\rho\pi\gamma} = 0.647$ and $g_\rho = 4.96$). 
At  $s=0$ the sum of the two resonance contributions to $F_{\pi\gamma}$ 
 agree with the anomaly driven normalization to within 10-15\% 
  and the isovector contribution can be further improved 
   using a unitary parametrization of  \cite{truong_fpi,truong_3piga}, which 
    for $F_{2\pi}$  and $t_{2\pi,\pi\gamma}$ in  Eq.~(\ref{imt}) yields, 
   \begin{equation} 
   F_{2\pi}(s)   =   P(s)  \Omega(s), \;\;   t_{2\pi,\pi\gamma}(s) =
   C^{-1}(s) \Omega(s), \label{rho}
   \end{equation}
  where  $\Omega(s)$, the Omnes-Muskelishvilli function \cite{omnes} computed  from the phase of the 
   vector-isovector  elastic $\pi\pi$ scattering amplitude and satisfying the VMD relation, 
 $ \Omega(s \sim m_\rho^2) \sim  m^2_\rho/(m_\rho^2 - s -i m_\rho \Gamma_\rho(s))$. The  polynomials  $P(s)$ and $C(s)$ ($P(s)=1+0.1 s/m_\rho^2  $, $C(s) = f^2_\pi[1 + 1.27
      s/m^2_\rho   +1.38 s^2/m^4_\rho  - 0.50 s^3/m^6_\rho] $) 
 are  determined from fits to the electromagnetic form factor and the
solution to the dispersion relation  for the $t_{2\pi,\pi\gamma}$
amplitude, respectively.  At higher energies, $s \gsim 1{ GeV}^2$ the $K{\bar K}$ 
inelastic channel 
and other multi-particle intermediate states are expected to contribute. 
Unfortunately, since no time-like data are available (unlike 
in the case of the electromagnetic form factor) one cannot 
unambiguously determine these contributions. 
A possible determination of the multi-particle hadronic  states could 
 be given in terms of quark/gluon intermediate states, much 
like in the derivation of QCD sum rules ({\it cf.} Ref.~\cite{qcd_sumrules} for the case of the 
pion electromagnetic form factor). 

\begin{figure}[h]
\centering
\vspace{-0.5cm}
\includegraphics[width=9.5cm]{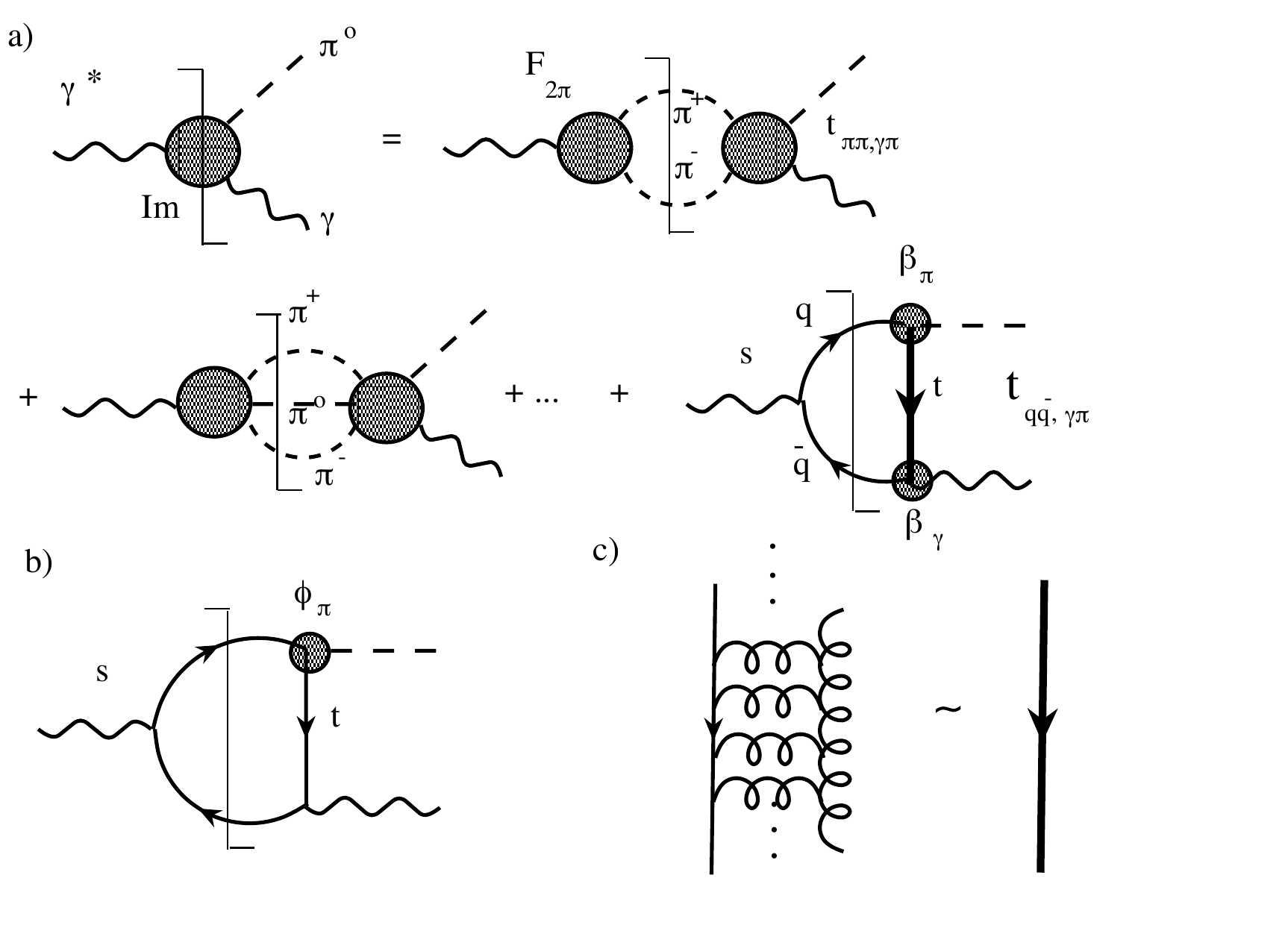}
\caption{Hadronic and asymptotic contributions to the $\pi^0$
  transition form factor.}
\label{fig1}
\end{figure}

\begin{figure}[h]
\vspace{-0.5cm}
\includegraphics[width=9cm,trim=0 40 0 -20]{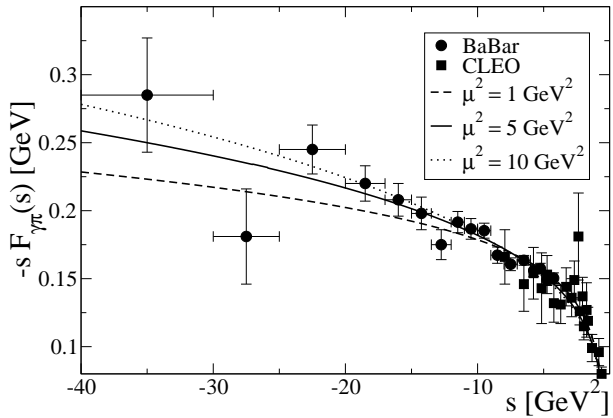}
\caption{Our results for $|F_{\pi\gamma}(s)|$ in the
  space-like region for  $\mu^2=$ 1 GeV$^2$ (dashed line), 
5 GeV$^2$ (solid line), 10 GeV$^2$ (dotted line),  in comparison with
the experimental data from \cite{babar,cleo}. } 
\label{fitt}
\end{figure}

Since the electromagnetic 
form factor $F_X$ of a composite state decreases with energy-momentum transfer,    
asymptotically the {\it r.h.s}  of Eq.~(\ref{imt}) is dominated 
by the $X = q \bar q$, quark-antiquark intermediate state. Its 
form factor is $F_{q\bar q}=1$, (in units of the quark charge)  and the state contributes to 
$Im F_{\pi \gamma}$ via the $q \bar q \to \pi \gamma$, 
$P$-wave  scattering amplitude, $t_{q\bar q,\pi\gamma}$ as illustrated  
by the last diagram in Fig.\ref{fig1}a.  The $q\bar q$ contribution 
shown in  Fig.\ref{fig1}a may be compared to the one in  Fig.\ref{fig1}b,  which 
represents the  asymptotic contribution as predicted by pQCD.   In the 
latter, the $q\bar q \to \pi\gamma$ scattering amplitude, shown to the
right of the vertical cut line, is given by a free quark propagator 
exchanged between the final state pion and photon.  In 
the kinematically relevant  domain  $s\gg t \sim b^{-1} $ with $t$
 being the four momentum squared carried by the exchanged quark  and 
$b \approx \mbox{ few GeV}^{-2}$ the typical slope of the product of 
residual coupling of the exchange quark ($\beta_\pi,\beta_\gamma$),  
the amplitude $t_{q\bar q,\pi\gamma}$ is expected to have a Regge
behavior \cite{fadin_sherman}
 \begin{equation} 
 t_{q\bar q,\pi\gamma}(s,t)  = \beta_\pi(t) \beta_\gamma(t) s^{\alpha_q(t)} \approx 
  e^{ b t } s^{\alpha_q}  \label{regge} 
 \end{equation} 
The difference between the free, Fig.\ref{fig1}b and the Regge propagator  
Fig.\ref{fig1}a can originate 
from the sum of ladder gluons in the wee region ({\it cf.} Fig.\ref{fig1}c).   
The quark Regge trajectory $\alpha_q(t)\approx\alpha_q(0)+\alpha'_qt$ 
is not known, however, 
phenomenologically it can be related to the leading Regge exchange  in $\pi\pi$ scattering, {\it i.e.} 
the $\rho$ or $f_2$ exchange (that are nearly degenerate, {\it cf.}~\cite{pelaez}),
$\alpha_\rho(t)\sim\alpha_{f_2}(t)$, 
\begin{equation} 
\alpha_q(t) \sim  0.5 \alpha_\rho(t) + 0.5   \approx 0.75+0.45 t. \label{a}
 \end{equation} 
It is worth noting that the dominance of quark-exchange (or, more
precisely, quark {\it interchange}) mechanism has previously 
been observed in hard scattering processes with hadrons~\cite{white}. 
Hard scattering data are furthermore compatible with
$\alpha_q(t=-1\,{\rm GeV}^2)\approx0.3-0.4$ \cite{strikman}, which is consistent with Eq.~(\ref{a}). 
Detailed  derivation of 
   Eq.~(\ref{a})  {\it i.e.} relation between quark and meson Regge trajectories  will be given in the 
   forthcoming paper~\cite{paper}.
After projecting onto spin-1 partial wave, the energy dependence 
of the asymptotic, $q\bar q$ contribution to $Im F_{\gamma\pi}$ is 
therefore expected to behave as (modulo terms $\sim O(\ln s)$),
    \begin{equation} 
    ImF^{(q\bar q)}_{\gamma^*\pi\gamma}(s) \to  c^{(q\bar q)}_{2\pi}  s^{\alpha_q(0) - 3/2}. 
       \label{ass} 
     \end{equation}  
 The important point is that with $\alpha_q(0) = 
 1/2+\epsilon$ (assuming $\alpha_q'(0) > 0$, 
it is consistent with the absence of a physical pole for a confined
quark) Eq.~(\ref{ass}) implies asymptotic increase of the energy
weighted transition 
form factor, $s F_{\pi\gamma}(s) \propto s^\epsilon$ in agreement 
with  the BaBar measurement. Such an increase cannot be accounted for by the 
exchange of the free quark as it is the case for the leading twist 
pQCD model. Combining the $\omega$ 
and the $\rho$ resonance contributions of Eqs.~(\ref{omega}),(\ref{rho})  with 
the asymptotic  form of Eq.~(\ref{ass}) and making the  simplifying 
assumption that the first two contribute to $Im F_{\pi\gamma}$ 
for $s< 1\mbox{ GeV}^2$ while the asymptotic part saturates 
$Im F_{\pi\gamma}$ for $s > \mu^2 $, we fit the the available data 
using Eq.~(\ref{disp}) with the single free parameter $c^{(q\bar q)}_{\pi\gamma}$ 
 that  determines the normalization of the asymptotic contribution.  
The result is shown in Fig.\ref{fitt}. It is worth noting that even
at largest values of $-s$ the bare $q\bar q$ production gives only
about $50\%$ of the form factor with the remaining half 
coming from the resonances.

In the case of the pion electromagnetic form factor, the discontinuity across the unitary cut is given by 
\begin{equation} 
Im F_{2\pi} = t^*_{2\pi,2\pi} \rho_{2\pi} F_{2\pi} +  t^*_{K\bar K,2\pi} \rho_{2K}  F_{K} + 
\sum_{X} t^*_{X,2\pi} \rho_X F_X  \label{uem}
\end{equation} 
where in the sum is over 
 intermediate states ($X\ne 2\pi, K\bar K$)  in 
$\gamma^* \to X \to 2\pi$ and where we separated 
 the two channels  $X=2\pi$ and $X=K\bar K$  
which phenomenologically are most significant in the hadronic 
domain~\cite{norma}.  Above the inelastic threshold,  $s> s_{i}$, the unitarity relation now 
involves both  $Im F_{2\pi}$ and $Re F_{2\pi}$ and can be solved 
algebraically. Assuming that the elastic amplitude, $t_{2\pi,2\pi}$ 
asymptotically  approaches the diffractive limit, 
$t_{2\pi,2\pi} \to i/2\rho_{2\pi}$, from Eq.~(\ref{uem}) one finds 
 \begin{equation} 
 F_{2\pi}(s) \to 2 i \sum_{X \ne 2\pi} t_{X,2\pi} \rho_X F^*_X \to 2 i t_{q\bar q,2\pi} \propto i s^{\alpha_q(0)-3/2}. 
 \end{equation} 
Except for the expected energy dependence {\it cf.} Eq.~(\ref{regge}), we do not know separately the real and  imaginary parts of $t_{q\bar q,2\pi}$. Assuming, as in the case 
of the transition form factor, that the real part of the 
discontinuity due to $q\bar q$  state has the energy dependence 
given by the reggized quark exchange, we can compute $F_{2\pi}$ using
Eq.~(\ref{uem}) and the Cauchy representation 
(the imaginary part of $t^*_{q\bar q,2\pi} \rho_X$ would then be given
by the solution of an algebraic equation that follows from
Eq.~(\ref{uem})). 
This yields, 
$F_{2\pi}(s) = N(s)/D(s)$ with 
 \begin{eqnarray} 
N(s)  &= &    \sum_{X\ne 2\pi} \frac{1}{\pi} \int_{s_i} ds' \frac{  
  D(s')  Re\left[t^*_{X,2\pi}(s') \rho_X(s') F_X(s')\right]}{ [1 - i t^*_{2\pi,2\pi}(s') \rho_{2\pi}(s')] (s' - s)} 
  \nonumber \\
  D(s) &  = &  \exp\left( -\frac{s}{\pi}  \int_{s_{th}} ds' \frac{ \phi(s')}{(s' -s)s'} \right).  \label{nd} 
  \end{eqnarray} 
 The phase $\phi$ is obtained from the elastic amplitude, 
$\tan\phi = {\rm Re} t_{2\pi,2\pi} \rho_{2\pi} /(1 - {\rm Im} t_{2\pi.2\pi}\rho_{2\pi})$. 
 As discussed earlier, the dominant feature of the pion electromagnetic 
form factor is the $\rho(770)$ resonance. Close to the resonance peak  there is
also a contribution  from the isospin-violating  $\omega \to 2\pi$ 
decay. Here we do not attempt to reproduce the details of the $\rho-\omega$ interference region. 
 The next relevant feature is the large variation in 
magnitude of $|F_{2\pi}|$ at $\sqrt{s} \sim 1.7 \mbox{ GeV}$ which is also seen in 
the elastic $2\pi \to 2\pi$ amplitude and is attributed to the
contribution  from the inelastic $\rho''(1700)$ resonance decaying to $K{\bar K}$.  
We thus approximate the sum over inelastic channels in Eq.~(\ref{nd}) by the single 
 $K{\bar K}$ channel,  and  above $s \ge \mu^2$ the residual $q\bar q$ continuum with 
\begin{equation} 
 Re t^*_{q\bar q,2\pi} \rho_X = c^{(q\bar q)}_{2\pi}  s^{\alpha_q(0)-3/2}. \label{asem} 
\end{equation} 
For the $t_{2\pi,2\pi}$ and $t_{K\bar K,2\pi}$ amplitudes we use the parametrization
 from ~\cite{peng}. Even though the contribution to the dispersive integral
 from the high energy tails of $t_{2\pi,2\pi}$ and $t_{K\bar K,2\pi}$
 are suppressed by form factors {\it cf.} Eq.~(\ref{uem}) we 
nevertheless extend  the parametrization from ~\cite{peng} to 
higher energies by smoothly joining  the resonance region to 
the  spin-1 projected  Regge limit of $\pi\pi\to \pi\pi$ and 
$K{\bar K} \to \pi\pi$ amplitudes.  
We parametrize the isovector kaon form factor $F_{K}$ using 
Breit-Wigner distributions  which include the  $\rho(770)$, 
$\rho'(1400)$ and  $\rho''(1700)$ \cite{fK}.  Finally we fit the available 
data on $|F_{2\pi}(s)|^2$ (excluding the $\rho-\omega$  
interference   region)  with five parameters: the magnitude and 
phase of the $\rho'$ and $\rho''$ contributions to $F_{K}$ 
and $c^{(q\bar q)}_{2\pi}$ --the magnitude of the $q\bar q$ continuum, 
Eq.~(\ref{asem}). 
\begin{figure}
\vspace{-0.5cm}
\includegraphics[width=9.5cm,trim=30 40 0 -20]{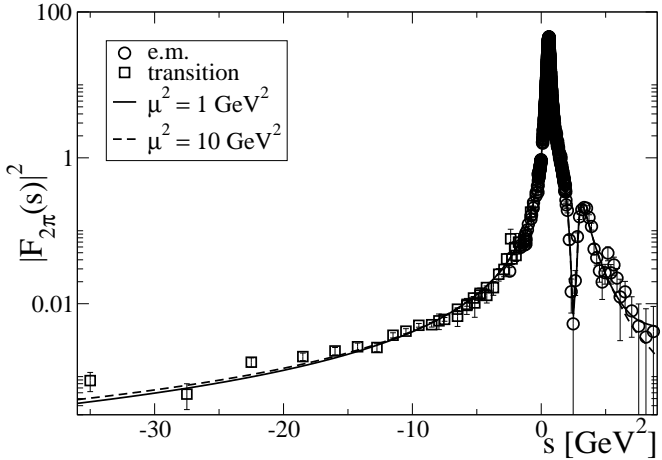}
\caption{Our results for the pion electromagnetic form factor for $\mu^2=1$ GeV$^2$
  (solid line) and $\mu^2=10$ GeV$^2$ (dashed line) vs. experimental
  data on the time-like and space-like e.-m. form factor from \cite{babar_em}
  (solid circles). }
\label{pionfit}
\end{figure}
 In Fig.~\ref{pionfit}, we display our results for the electromagnetic pion form 
factor $F_{\pi}$ in the range -40 GeV$^2\leq s\leq10$ GeV$^2$. We
confront them with the available experimental data for the
electromagnetic form factor for -10 GeV$^2\leq s\leq10$ GeV$^2$ and
the transition form factor for -40 GeV$^2\leq s\leq-0.8$ GeV$^2$
(both are normalized to 1 at $s=0$). First, we note that in the
space-like region the data sets for the two form factors look
identical (this is not expected to be the case for the time-like region since, for example the
$\omega(782)$ only contributes to the transition form factor). 
One can see that our model describes all the
available data throughout the shown kinematics. This serves as an {\it a
  posteriori} evidence that this $s$-dependence is in both cases
driven by the same mechanism. In the case of the electromagnetic form
factor, our result is a prediction for the $s$-dependence at large
$|s|$, where no data exist so far. In particular, we predict that, as 
for the transition form factor case, $|s\,F_{2\pi(s)}|$ has to rise
asymptotically roughly as $s^{1/4}$, unlike pQCD predictions that
feature at most a logarithmic limit for that combination. 

To summarize, we presented a calculation of  the transition and
electromagnetic form factors of the pion. We used dispersion relations
to provide a unified description of the pion form factors in the 
time-like and space-like regions. In the hadronic
energy range, we accounted for hadronic, resonance mechanisms in
a fully unitarized manner. For asymptotic contributions, we proposed a
new mechanism that features a reggeized quark exchange. 
We relate the parameters of such an exchange to $\pi\pi$ scattering data and deduce that the quark-Regge
intercept is  approximately  $\alpha_q(0) \sim 3/4$. Using this value as input for the
asymptotic behavior of the pion form factors, we obtain for both 
$s F(s)\propto s^{1/4}$, in agreement with the recent BaBar data. We notice that  when the transition form factor is renormalized so that $F_{\pi\gamma}(0)=1$  its dependence on $s$ in the space-like region is consistent with that of the electromagnetic form factor, as shown by the open circles in Fig.\ref{pionfit}. 
We use the normalization of the Regge-behaved 
$t_{q\bar q,2\pi}$ and $t_{q\bar q,\pi\gamma}$ as the only free
parameter, and are able to describe all available data on pion form
factors.   

 This work was supported in part by the US
Department of Energy under contract DE-FG0287ER40365 and the
National Science Foundation grants PHY-0555232 (M.G.), PIF-0653405 (P.G.).
\def\etal{\textit{et al.}}

\end{document}